%% file: Main.tex
\documentclass[pdflatex,sn-mathphys-num]{sn-jnl}

\usepackage[normalem]{ulem}
\usepackage{upgreek}
\usepackage{graphicx}%
\usepackage{multirow}%
\usepackage{amsmath,amssymb,amsfonts}%
\usepackage{amsthm}%
\usepackage{mathrsfs}%
\usepackage[title]{appendix}%
\usepackage{xcolor}%
\usepackage{textcomp}%
\usepackage{manyfoot}%
\usepackage{booktabs}%
\usepackage{algorithm}%
\usepackage{algorithmicx}%
\usepackage{algpseudocode}%
\usepackage{listings}%
\usepackage{bm}
\usepackage{pdfpages}

\theoremstyle{thmstyleone}%
\theoremstyle{thmstyletwo}%

\theoremstyle{thmstylethree}%

\begin{document}

\title[Article Title]{Local information processing capacity: anatomy of computation in neuromorphic systems}


\author*[1]{\fnm{Jiaxuan} \sur{Chen}}\email{CHEN.Jiaxuan@nims.go.jp}

\author[1]{\fnm{Daiki} \sur{Nishioka}}\email{NISHIOKA.Daiki@nims.go.jp}

\author[2]{\fnm{Yicheng} \sur{Song}}\email{song.yicheng@nitech.ac.jp}

\author[1]{\fnm{Ryo} \sur{Iguchi}}\email{IGUCHI.Ryo@nims.go.jp}

\author[1,3]{\fnm{Sota} \sur{Hikasa}}\email{HIKASA.Sota@nims.go.jp}

\author[1,3]{\fnm{Hina} \sur{Kitano}}\email{KITANO.Hina@nims.go.jp}

\author[4]{\fnm{Akira} \sur{Hirose}}\email{ahirose@ee.t.u-tokyo.ac.jp}

\author[1,3]{\fnm{Takashi} \sur{Tsuchiya}}\email{TSUCHIYA.Takashi@nims.go.jp}

\affil*[1]{\orgdiv{Research Center for Materials Nanoarchitectonics (MANA)}, 
\orgname{National Institute for Materials Science (NIMS)}, 
\orgaddress{\street{1-1 Namiki}, \city{Tsukuba}, \state{Ibaraki}, \postcode{305-0044}, \country{Japan}}}

\affil[2]{\orgdiv{Graduate School of Engineering}, \orgname{Nagoya Institute of Technology}, \orgaddress{\street{Gokiso-cho, Showa-ku}, 
\city{Nagoya}, 
\state{Aichi-ken}, 
\postcode{466-8555}, 
\country{Japan}}}

\affil[3]{\orgdiv{Faculty of Advanced Engineering}, \orgname{Tokyo University of Science}, \orgaddress{\street{6-3-1 Niijuku}, \city{Katsushika}, \state{Tokyo}, \postcode{125-8585}, \country{Japan}}}

\affil[4]{\orgdiv{Department of Electrical Engineering and Information Systems, Graduate School of Engineering}, 
\orgname{The University of Tokyo}, 
\orgaddress{\street{7-3-1 Hongo}, 
\city{Bunkyo-ku}, 
\state{Tokyo}, 
\postcode{113-8656}, 
\country{Japan}}}


\abstract{Many neuromorphic systems possess rich internal computational structures shaped by their physical dynamics. However, existing metrics characterize only global computation, leaving these local structures unresolved despite their importance for system interpretation, optimization and miniaturization. Here, we introduce local information processing capacity (L-IPC), which represents individual dynamical states as vectors in an infinite-dimensional Hilbert space and decomposes their computational properties through subspace projections onto nonlinearity orders and memory steps, with the residual quantifying computational irrelevance. Applied to three systems with comparable global capacities --- an echo state network, an ion-gating reservoir and a spin-wave reservoir --- L-IPC reveals markedly different distributions of computation, ranging from homogeneous to spatially, temporally and spectrally structured patterns. These local profiles further guide task-specific pruning, outperforming standard approaches at the same size in the structured systems. These results establish L-IPC as a general framework for anatomizing computation and guiding task-adaptive design in neuromorphic systems.}




\maketitle

\section*{Introduction}

Dynamical systems have long been recognized as powerful resources for computation~\cite{maass2002real,jaeger2001echo}. Neuromorphic computing builds on this principle, harnessing the intrinsic dynamics of these systems --- from spiking neural networks~\cite{roy2019towards,PhysRevLett.125.207202,feldmann2019all,dang2026spiking} to physical reservoir computing~\cite{tanaka2019recent,yan2024emerging,PhysRevX.10.041037,lee2024task,zhong2022memristor} --- to overcome the energy and latency limitations of conventional von Neumann architectures~\cite{markovic2020physics,wright2022deep,kurebayashi2026metrics}. At the same time, the complexity and diversity of these dynamics make such systems difficult to interpret, rendering them black boxes in practice.

\begin{figure}
\centering
\includegraphics[width=0.88\textwidth]{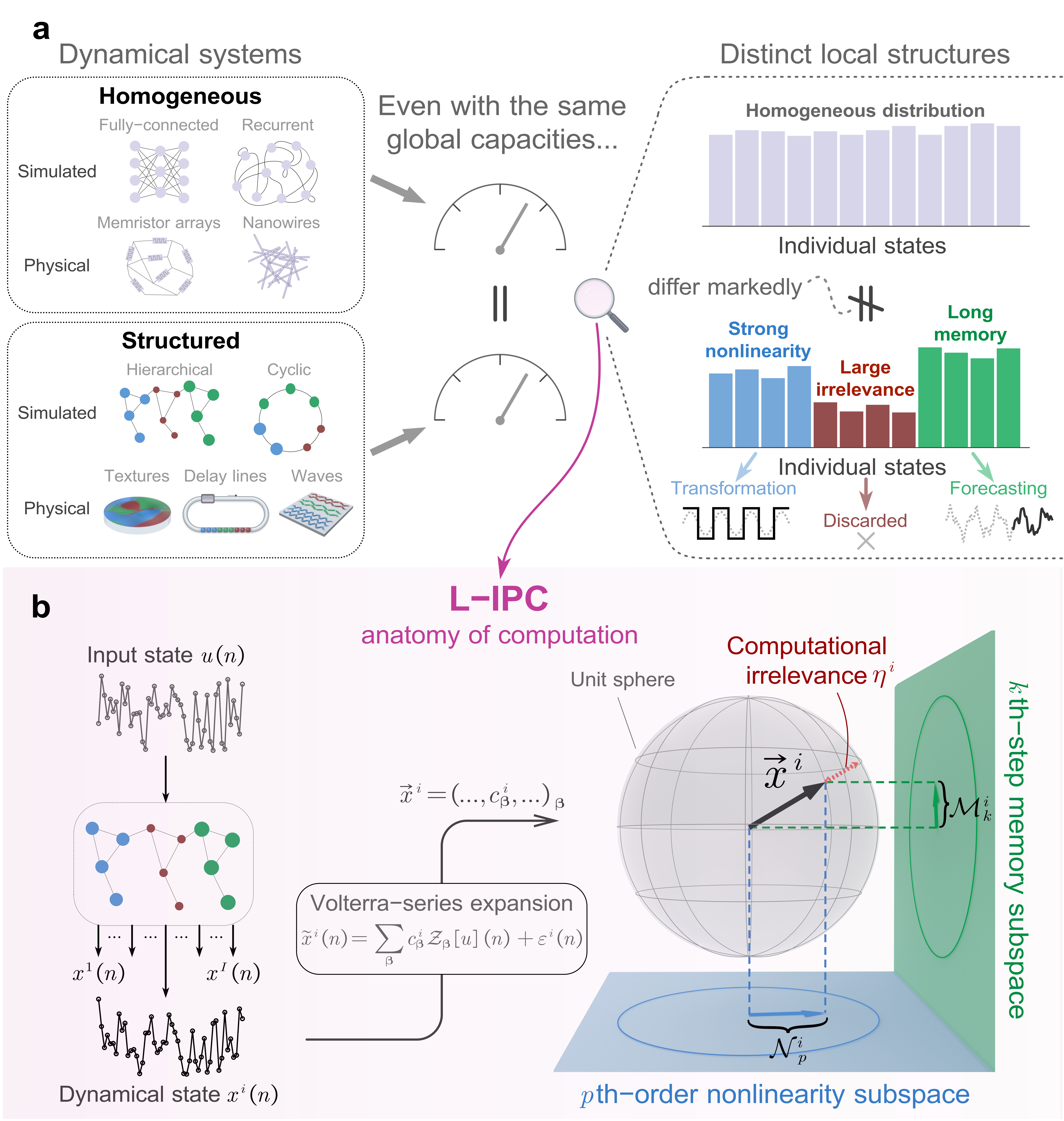}
\caption{\textbf{\textbar\ L-IPC reveals the anatomy of computation within a system.} \textbf{a}, Examples of homogeneous and structured dynamical systems, both simulated and physical (left). Even when the two types of systems have equal global capacities (middle), their local structures markedly differ (right): the homogeneous system distributes capacity uniformly across states, whereas in the structured system the states separate into groups dominated by strong nonlinearity, long memory or large computational irrelevance. Resolving these roles enables task-specific state selection, retaining the states a task requires and discarding the irrelevant ones. \textbf{b}, Conceptual illustration of L-IPC. An input $u(n)$ drives the system, producing dynamical states $x^1(n),\ldots,x^I(n)$. Each normalized state $\tilde{x}^i(n)$ is expanded in a Volterra series into a vector $\vec{x}^{\,i}$ in a Hilbert space, together with an input-independent residual $\varepsilon^i(n)$ outside that space. Projecting $\vec{x}^{\,i}$ onto functional subspaces yields the $p$th-order nonlinearity $\mathcal{N}_p^i$ (blue) and the $k$th-step memory $\mathcal{M}_k^i$ (green). The gap $\eta^i$ between the tip of $\vec{x}^{\,i}$ and the unit sphere quantifies the computational irrelevance arising from $\varepsilon^i(n)$ (dark red).}
\label{fig:localandglobal}
\end{figure}

Several metrics have been developed to characterize the global computational capabilities of such systems, including memory capacity~\cite{jaeger2001short}, kernel and generalization ranks~\cite{legenstein2007edge,dale2019substrate}, and Lyapunov exponents~\cite{inubushi2017reservoir}. Among these, information processing capacity (IPC) stands out as a unified and task-independent measure~\cite{dambre2012information}, quantifying the ability of the system as a whole to reconstruct functions of past inputs across nonlinearity orders and memory steps~\cite{vettelschoss2021information,tsunegi2023information,martinez2023information,hikasa2026enhanced}. However, these global metrics capture only how much a system can compute, providing limited insight into how computation is organized across the individual dynamical states, i.e., the time-sequential response signals obtained from the system and used to construct the output. In particular, systems with comparable global capacities can differ markedly in their distribution of computation, revealing a fundamental distinction between computationally homogeneous and structured systems (Fig.~\ref{fig:localandglobal}a).

Understanding what individual dynamical states compute would provide an anatomy of computation within a material or device, identifying where computation resides, which resources are missing or underutilized, and which states are required for a given task. Analyses of this kind have already proved their value in artificial neural networks: determining what individual neurons compute has revealed that only a small subset is critical for a particular task~\cite{bau2020understanding} and has enabled pruning based on component importance~\cite{molchanov2017pruning}. The need is more acute in neuromorphic systems, where each state must be physically read out, making a large number of states costly in device size, complexity and training overhead~\cite{yan2024emerging,kurebayashi2026metrics}. Yet what these states contribute is not obvious, as they arise from physical dynamics whose intricacy can also produce computationally irrelevant responses. A general framework for anatomizing computation in neuromorphic systems therefore remains to be established.

Here, we introduce local information processing capacity (L-IPC), a framework for resolving the computational structure of individual dynamical states. L-IPC represents each state as a vector in an infinite-dimensional Hilbert space and characterizes its computational role through projections onto subspaces of specific nonlinearity orders and memory steps, with the residual quantifying its computational irrelevance (Fig.~\ref{fig:localandglobal}b). Although capacity is conventionally defined for a system as a whole, we show that it remains meaningful in a local sense: the distribution of capacities across these subspaces reflects the underlying dynamical and physical organization of the system. In particular, the computational irrelevance quantified by L-IPC is an intrinsically local property that has remained inaccessible to global measures. To show the system-agnostic nature of L-IPC, we apply it to three dynamical systems --- an echo state network (ESN) and two physical reservoirs, one based on ion-gating dynamics with temporally resolved states~\cite{nishioka2025two} and the other on spin-wave dynamics with spectrally resolved states~\cite{chen2026spectral} --- and uncover distinct computational structures despite their similar global capacities. We further demonstrate that these local profiles enable task-adaptive state selection in computationally structured systems, substantially reducing the system size while achieving lower prediction errors than random selection or selection by readout weight at the same number of states.

\section*{Results}

\subsection*{Formulation of L-IPC}

L-IPC is grounded in the universality of the Volterra-series approximation for fading-memory dynamical systems~\cite{boyd1985fading,grigoryeva2019differentiable}. As illustrated in Fig.~\ref{fig:localandglobal}b, consider a dynamical system driven by an input $u(n)$ that produces $I$ dynamical states $x^1(n),\ldots,x^I(n)$. Each state is normalized as $\tilde{x}^i(n) = (x^i(n) - \mu_i)/\sigma_i$, where $\mu_i$ and $\sigma_i$ are its mean and standard deviation, and is then expanded as
\begin{equation}
    \tilde{x}^i(n) = \sum_{p=1}^{\infty} \sum_{\bm{\upbeta}:\,|\bm{\upbeta}|=p} c^i_{\bm{\upbeta}}\, \mathcal{Z}_{\bm{\upbeta}}[u](n) + \varepsilon^i(n).
    \label{eq:volterra}
\end{equation}
Here, the first term $\sum_{\bm{\upbeta}} c^i_{\bm{\upbeta}}\, \mathcal{Z}_{\bm{\upbeta}}[u](n)$ represents the input-driven part of the state, where $\bm{\upbeta} = (\beta_0, \beta_1, \ldots)$ is a multi-index whose entry $\beta_k$ is the polynomial degree applied to the input at delay step $k$, and $|\bm{\upbeta}| = \sum_k \beta_k$ is the nonlinearity order. The functionals $\{\mathcal{Z}_{\bm{\upbeta}}[u]\}_{\bm{\upbeta}}$ form an orthonormal basis with respect to the input distribution (Methods) and $\{c^i_{\bm{\upbeta}}\}_{\bm{\upbeta}}$ are the expansion coefficients. The second term $\varepsilon^i(n)$ is the residual that collects the input-independent part of the state, which enables us to quantify a state's computational irrelevance, something global measures such as IPC cannot resolve.

These coefficients $\{c^i_{\bm{\upbeta}}\}_{\bm{\upbeta}}$ directly encode the computational properties of each state, as a mathematical equivalence was established between their squares and IPC~\cite{PhysRevResearch.3.043135}. It is therefore natural to identify each state with a vector $\vec{x}^{\,i} = (\ldots, c^i_{\bm{\upbeta}}, \ldots)_{\bm{\upbeta}}$ in the Hilbert space $\mathcal{H}$ spanned by the basis functionals (Fig.~\ref{fig:localandglobal}b). The direction of $\vec{x}^{\,i}$ specifies what the state computes, through its projections onto two families of functional subspaces, the nonlinearity subspace of order $p$, $\mathcal{H}^{\mathrm{NL}}_{p} = \mathrm{span}\{\mathcal{Z}_{\bm{\upbeta}}[u] : |\bm{\upbeta}| = p\}$, and the memory subspace of step $k$, $\mathcal{H}_k^{\mathrm{MEM}} = \mathrm{span}\{\mathcal{Z}_{\bm{\upbeta}}[u] : \beta_k \neq 0\}$. Correspondingly, we define the $p$th-order nonlinearity $\mathcal{N}_p^i$ and the $k$th-step memory $\mathcal{M}_k^i$ of state $x^i(n)$ as the lengths of these projections
\begin{align}
    \mathcal{N}_p^i &= \|P_p^{\mathrm{NL}}\, \vec{x}^{\,i}\| = \sqrt{\sum_{|\bm{\upbeta}|=p} (c^i_{\bm{\upbeta}})^2}, \label{eq:Np} \\
    \mathcal{M}_k^i &= \|P_k^{\mathrm{MEM}}\, \vec{x}^{\,i}\| = \sqrt{\sum_{\bm{\upbeta}:\,\beta_k \neq 0} (c^i_{\bm{\upbeta}})^2}, \label{eq:Mk}
\end{align}
where $P_p^{\mathrm{NL}}$ and $P_k^{\mathrm{MEM}}$ are the orthogonal projection operators onto these subspaces, and $\|\cdot\|$ is the norm induced by the time-averaged inner product (Methods). By the equivalence above, these quantities rest on the same theoretical foundation as IPC while resolving computation at the level of individual states.

The residual $\varepsilon^i(n)$ in the expansion~\eqref{eq:volterra} isolates the state's computational irrelevance, which can arise from noise, environmental disturbance or chaotic behavior. Such isolation is possible only in the local sense, as any global measure would conflate this irrelevance with the redundancy among states. Geometrically, a state entirely determined by the input history places $\vec{x}^{\,i}$ on the unit sphere of $\mathcal{H}$, whereas a nonzero $\varepsilon^i(n)$ draws it inside. We define the computational irrelevance of $x^i(n)$ as the radial gap to that sphere,
\begin{equation}
    \eta^i = 1 - \|\vec{x}^{\,i}\| = 1 - \sqrt{\sum_{\bm{\upbeta}} (c^i_{\bm{\upbeta}})^2},
    \label{eq:eta}
\end{equation}
which ranges from $\eta^i = 0$ for a state with no residual to $\eta^i = 1$ for one that carries no information about the input.

Beyond $\eta^i$, the L-IPC expansion in Eq.~\eqref{eq:volterra} separates the irrelevant part of each state from its input-driven part directly as a time series, allowing us to quantify the capacity a system would provide in the absence of irrelevance. Comparing this with the capacity obtained from the raw states distinguishes what a system loses to irrelevance from what it loses to redundancy among states. This offers a new design perspective that a global capacity alone cannot provide, namely whether performance is better improved by suppressing noise and stabilizing the dynamics or by seeking richer nonlinearity and longer memory. Demonstrating this with a dynamically unstable ESN, we find that more than half of the available capacity is concealed by irrelevance, most severely at high nonlinearity orders and long memory steps (Supplementary Fig.~1).

Together, $\{\mathcal{N}_p^i\}_{p}$, $\{\mathcal{M}_k^i\}_{k}$ and $\eta^i$ constitute the L-IPC profile of state $x^i(n)$, an anatomy of its computation. We validated L-IPC on ESNs, whose dynamical properties are well understood~\cite{jaeger2001echo,lukovsevivcius2009reservoir}: the profiles track the nonlinearity and memory of individual states consistently with the global behavior of these networks, and $\eta^i$ rises with dynamical instability and with stochastic noise (Supplementary Note~1 and Supplementary Fig.~2).

\begin{figure}
\centering
\includegraphics[width=0.99\textwidth]{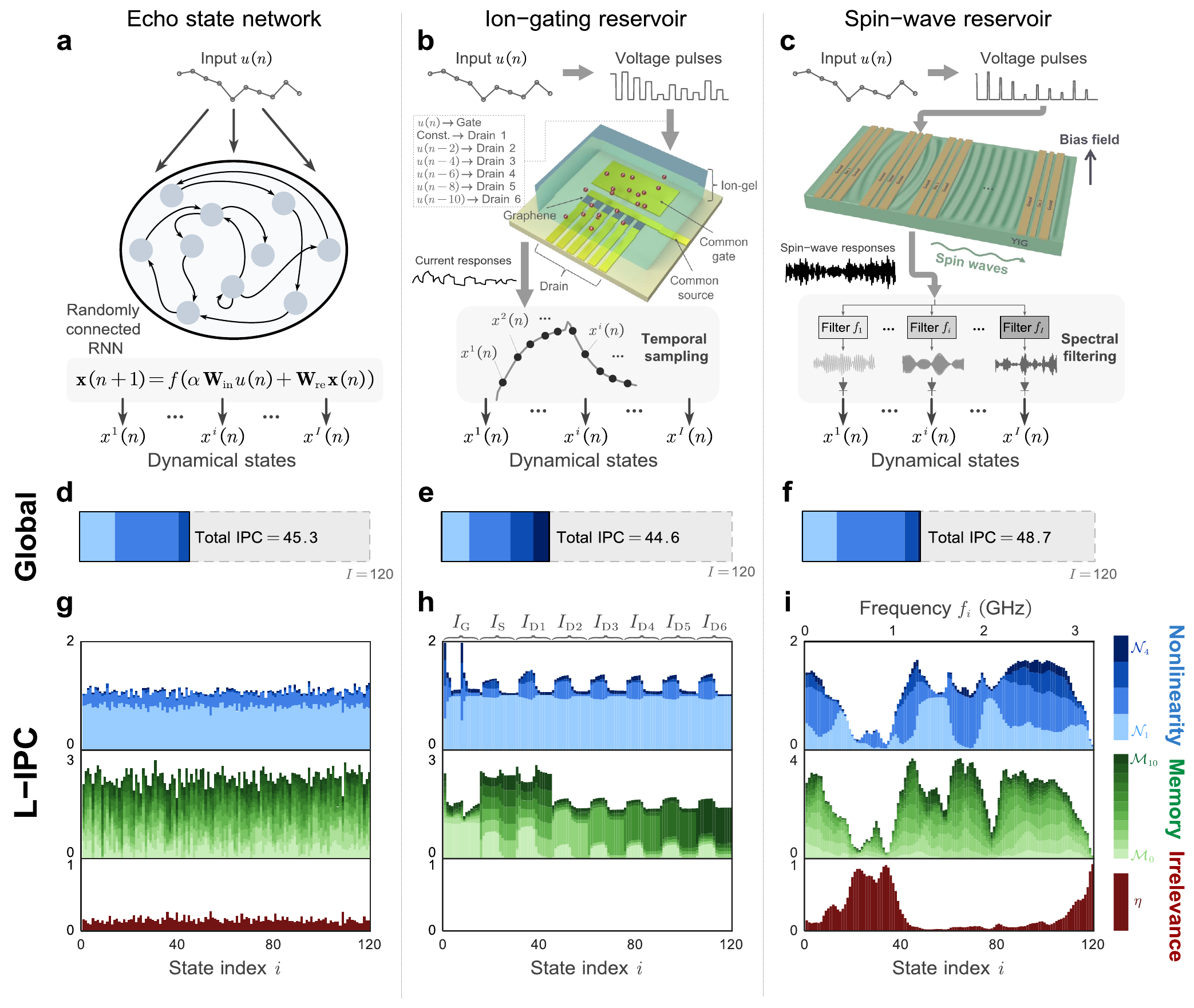}
\caption{\textbf{\textbar\ L-IPC reveals distinct computational structures in three systems of comparable global capacities.}
\textbf{a}--\textbf{c}, The three dynamical systems providing the same number of states, $I = 120$. \textbf{a}, An ESN, whose states are the discrete-time activations of a randomly connected recurrent neural network driven by the input $u(n)$. \textbf{b}, An IGR with a common gate, a common source and six drains, in which voltage pulses encoding $u(n)$ are applied to the gate to drive nonlinear ionic transport, while delayed copies or a constant bias are applied to the drains; the states are extracted from the current responses by temporal sampling within the pulse interval. \textbf{c}, An SWR, in which voltage pulses excite spin-wave propagation in a YIG single crystal and the response is measured at a detector CPW; the states are extracted by spectral filtering of the detected response at frequencies $f_i$, followed by envelope detection. \textbf{d}--\textbf{f}, Global IPC of the three systems, decomposed by nonlinearity order. The solid bar gives the total IPC and the dashed extension marks the upper bound $I = 120$. \textbf{g}--\textbf{i}, L-IPC profiles of all $120$ states: nonlinearity $\mathcal{N}_p^i$ for $p = 1$--$4$ (blue, top), memory $\mathcal{M}_k^i$ for $k = 0$--$10$ (green, middle) and computational irrelevance $\eta^i$ (red, bottom), with the components of each quantity stacked. In \textbf{h}, states are grouped by the electrode at which the current is measured ($I_\mathrm{G}$, $I_\mathrm{S}$ and $I_{\mathrm{D}1}$--$I_{\mathrm{D}6}$). In \textbf{i}, the upper axis gives the filter frequency $f_i$ of each state.}
\label{fig:interpretation}
\end{figure}

\subsection*{Distinct local structures at comparable global capacities}

We apply L-IPC to three dynamical systems, each with $I = 120$ states (Fig.~\ref{fig:interpretation}a--c). The first is an ESN~(Fig.~\ref{fig:interpretation}a), whose states are the discrete-time activations of a randomly connected recurrent neural network (Methods). The second is an ion-gating reservoir (IGR)~(Fig.~\ref{fig:interpretation}b), consisting of a graphene channel covered by ion gel, with a common gate, a common source and six drain electrodes. Ionic transport at the gel/graphene interface drives electric-double-layer formation and relaxation, providing the nonlinearity and memory for computation~\cite{nishioka2025two,nishioka2022edge}~(Supplementary Fig.~3). Its states are extracted through temporal sampling of the currents at all eight electrodes within each input pulse (virtual nodes~\cite{appeltant2011information}; Methods). The third is a spin-wave reservoir (SWR)~(Fig.~\ref{fig:interpretation}c), consisting of a yttrium iron garnet (YIG) single crystal patterned with coplanar waveguides (CPWs) under a bias magnetic field~\cite{chen2026spectral}, with nonlinear interaction and dispersive propagation of spin waves providing computationally rich dynamics~\cite{papp2021nanoscale,namiki2023experimental,korber2023pattern,chen2026nonlinear}. Its states are extracted through spectral filtering and envelope detection of the spin-wave responses (spectral nodes~\cite{chen2026spectral}; Methods). Thus, the three systems define their states through distinct architectural and physical features: network connectivity in the ESN, electrode location and sampling time in the IGR, and filter frequency in the SWR.

From a global perspective, the three systems are very similar in their computational capacities (Fig.~\ref{fig:interpretation}d--f), with total IPCs of $45.3$, $44.6$ and $48.7$, respectively, albeit with some differences in their decompositions by nonlinearity order. Through L-IPC, we find that their internal structures are markedly distinct. Figure~\ref{fig:interpretation}g--i shows the L-IPC profiles across states for nonlinearity $\mathcal{N}_p^i$ from $p = 1$ to $4$ (top), memory $\mathcal{M}_k^i$ from $k = 0$ to $10$ (middle), and computational irrelevance $\eta^i$ (bottom). The IGR states are grouped by the eight electrodes from which the currents are measured (Fig.~\ref{fig:interpretation}h), and the SWR states are additionally labeled by their filter frequency $f_i$ on the upper axis (Fig.~\ref{fig:interpretation}i). 

In the ESN, the L-IPC profiles are highly homogeneous across states (Fig.~\ref{fig:interpretation}g), consistent with the random connectivity of the network. The states are predominantly characterized by $\mathcal{N}_1^i$, with moderate $\mathcal{N}_2^i$ and negligible contributions from higher orders. All states exhibit memory across multiple delay steps, together with moderate and relatively uniform irrelevance $\eta^i$. These results suggest that the states of the ESN play similar computational roles, with little systematic differentiation across the network.

In contrast, the L-IPC profiles reveal the internal structure of the IGR as organized across both physical channels and sampling times (Fig.~\ref{fig:interpretation}h), knowledge that existing measures cannot access. Across channels, the gate current $I_\mathrm{G}$ carries the strongest nonlinearity, particularly in its first few sampling points. The memory profile follows the delay architecture of the device, with $I_{\mathrm{D}2}$--$I_{\mathrm{D}6}$ each carrying memory at the delay of the input copy applied to its drain (Methods). L-IPC also reveals that memory is not confined to those channels. The source $I_\mathrm{S}$ and first-drain $I_{\mathrm{D}1}$ currents, neither of which receives a delayed copy, still show moderate contributions across a range of memory steps $\mathcal{M}_k^i$. We attribute this to $I_\mathrm{S}$ integrating ionic transport from all drain channels and to $I_{\mathrm{D}1}$ picking up ionic crosstalk from its neighbors. Even within each channel, both the nonlinearity and memory profiles vary systematically with sampling time. Nonlinearity remains high while the pulse is on and peaks around the on--off transition of the input pulse, when the ionic response changes most rapidly. Instantaneous memory $\mathcal{M}_0^i$ is likewise substantial while the pulse is on and falls away when it is off. Computational irrelevance remains negligible throughout, indicating that the responses are almost entirely input-driven.

L-IPC shows that the SWR is also structured, but in a different pattern, with the profiles varying smoothly with filter frequency (Fig.~\ref{fig:interpretation}i). Computational capacity is distributed over a broad frequency range, reflecting the diverse spin-wave modes supported by the YIG single crystal (Supplementary Fig.~4). The higher-order nonlinear components ($\mathcal{N}_2^i$--$\mathcal{N}_4^i$) peak in several bands, around $f_i = 1.2$--$1.4$, $1.6$--$1.8$ and $2.2$--$2.7$~GHz, and fall into pronounced troughs between them, where the linear component $\mathcal{N}_1^i$ instead dominates. These peaks and troughs resemble the structure of the measured spin-wave transmission (Supplementary Fig.~4), suggesting a close link between the transmitted spin-wave power and the nonlinear computational capacity. The memory profile is concentrated in the same bands but extends over different numbers of delay steps, which we attribute to the dispersion and damping of the modes contributing at each frequency. Computational irrelevance becomes large outside the spin-wave band region ($1.2$--$2.7$~GHz), as background noise starts to dominate. However, the profiles also reveal appreciable capacity at the low-frequency end ($f_i < 0.5$~GHz), where the measured transmission shows no spin-wave band. This capacity may originate from direct electromagnetic crosstalk or from nonlinear mixing among spin-wave modes.

Taken together, these results show that L-IPC uncovers markedly different distributions of computation behind comparable global capacities, ranging from homogeneous to physically structured. By applying the same decomposition across three platforms with different state definitions, we establish L-IPC as a system-agnostic framework for anatomizing computation in neuromorphic systems.

\subsection*{Task-adaptive state selection}

\begin{figure}
\centering
\includegraphics[width=1\textwidth]{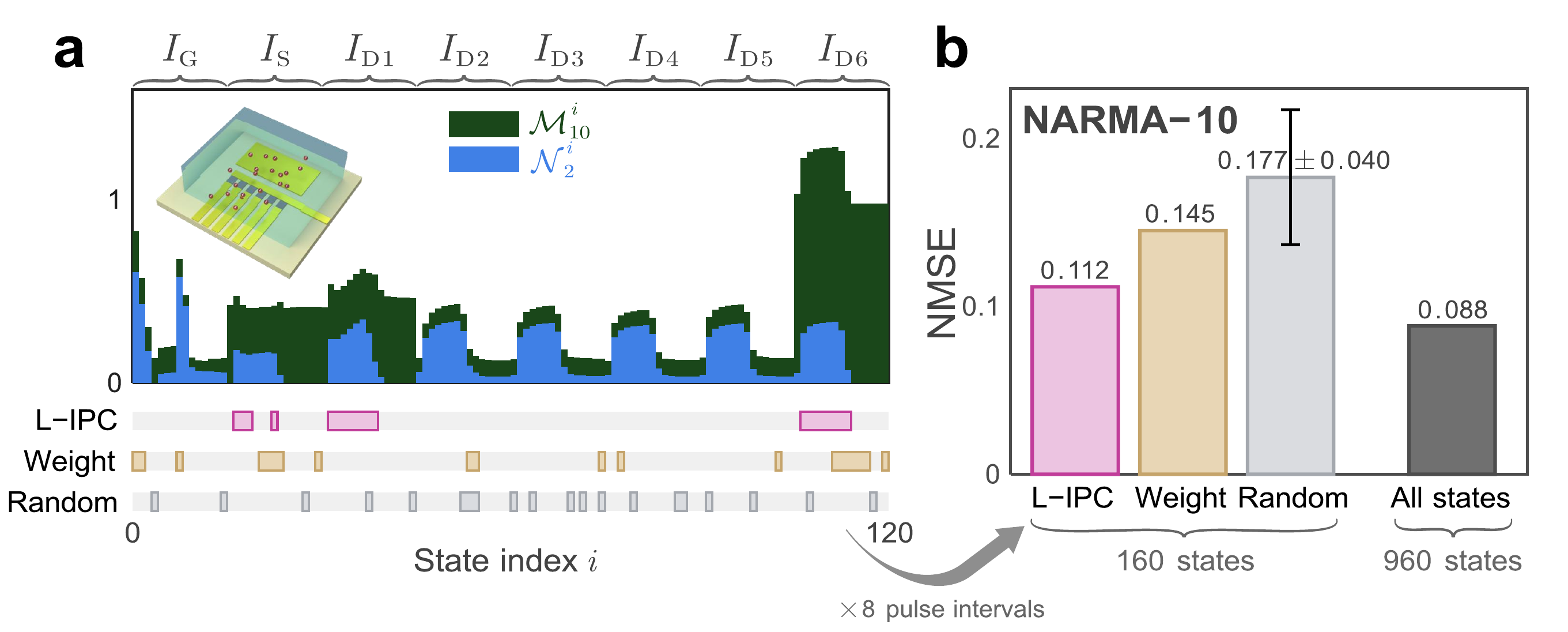}
\caption{\textbf{\textbar\ Comparison of L-IPC-guided state selection with baselines.} \textbf{a}, Task-relevant L-IPC components for NARMA-$10$, showing the second-order nonlinearity $\mathcal{N}_2^i$ (blue) and the tenth-step memory $\mathcal{M}_{10}^i$ (green), stacked, for the $120$ states of the IGR measured at a pulse interval of $80$~\textmu s, the same condition as in Fig.~\ref{fig:interpretation}h. The tracks below mark which of these states are selected by L-IPC (magenta), by readout weight magnitude (ochre) and at random (gray); in each case $160$ states are selected from the full pool of $960$ obtained by operating the IGR at eight pulse intervals. \textbf{b}, NARMA-$10$ NMSE of the three selections, each using $160$ states, together with the result obtained using all $960$ states. The error bar on the random selection is the standard deviation over $20$ independent draws.}
\label{fig:selection}
\end{figure}

Knowing what each state computes makes it possible to identify important states according to what a task demands. We first demonstrate this on the IGR, whose state set we expand to $I = 960$ by operating the device at eight pulse intervals (Methods). With a state set this large, selecting a subset becomes desirable under practical readout constraints~\cite{tanaka2019recent,kurebayashi2026metrics}. As a benchmark, we use the $l$th nonlinear autoregressive moving average (NARMA-$l$) task~\cite{atiya2000new,rodan2011minimum}, whose target combines a quadratic nonlinearity with a dependence on the input $l$ steps earlier (Methods). We therefore regard states with large second-order nonlinearity $\mathcal{N}_2^i$ and $l$th-step memory $\mathcal{M}_l^i$ as important for NARMA-$l$, and consider three tasks, $l = 6$, $8$ and $10$, to show task adaptivity. 

Figure~\ref{fig:selection}a shows the task-relevant components $\mathcal{N}_2^i$ and $\mathcal{M}_{10}^i$ for NARMA-$10$, extracted from the full L-IPC profile of the IGR operating at a pulse interval of $80$~\textmu s (Fig.~\ref{fig:interpretation}h). From these components, we select states to reduce the system size by first excluding those with high computational irrelevance $\eta^i$, whose input-unrelated responses can be detrimental to performance. Among the rest, we then identify the states with strong task-relevant memory $\mathcal{M}_l^i$ and finally retain those with large $\mathcal{N}_2^i$ (Methods). We compare this L-IPC-guided selection against two baselines: selection by readout weight magnitude, obtained after training on all $960$ normalized states, and random selection (Methods). As an example, we select $160$ states from the pool of $I = 960$. The tracks below the profile in Fig.~\ref{fig:selection}a mark the states chosen by each approach at this pulse interval. L-IPC-guided selection (magenta) returns blocks of states with large $\mathcal{M}_{10}^i$ and $\mathcal{N}_2^i$. The most prominent block lies in $I_{\mathrm{D}6}$, which receives the input delayed by ten steps, with further blocks in $I_{\mathrm{S}}$ and $I_{\mathrm{D}1}$, which carry substantial crosstalk. Within each channel, states sampled during the pulse-on phase are preferentially selected owing to their larger $\mathcal{N}_2^i$. On the other hand, weight-based selection (ochre) returns some states with large $\mathcal{M}_{10}^i$ and $\mathcal{N}_2^i$ but also many from regions where both components are small, without a clear pattern. Finally, random selection (gray) samples uniformly by construction. These differences translate directly into performance (Fig.~\ref{fig:selection}b). For NARMA-$10$ at $160$ states, L-IPC-guided selection yields a normalized mean square error (NMSE) of $0.112$, well below the $0.145$ for weight-based selection and $0.177 \pm 0.040$ for random selection at the same number of states, approaching the value obtained using all $960$ states ($0.088$). The L-IPC profiles and selected states for the other pulse intervals and NARMA tasks are provided in Supplementary Fig.~5, where the states selected by L-IPC adaptively shift to $I_{\mathrm{D}4}$ and $I_{\mathrm{D}5}$ for NARMA-$6$ and NARMA-$8$ in accordance with the delay architecture. 

\begin{figure}
\centering
\includegraphics[width=1\textwidth]{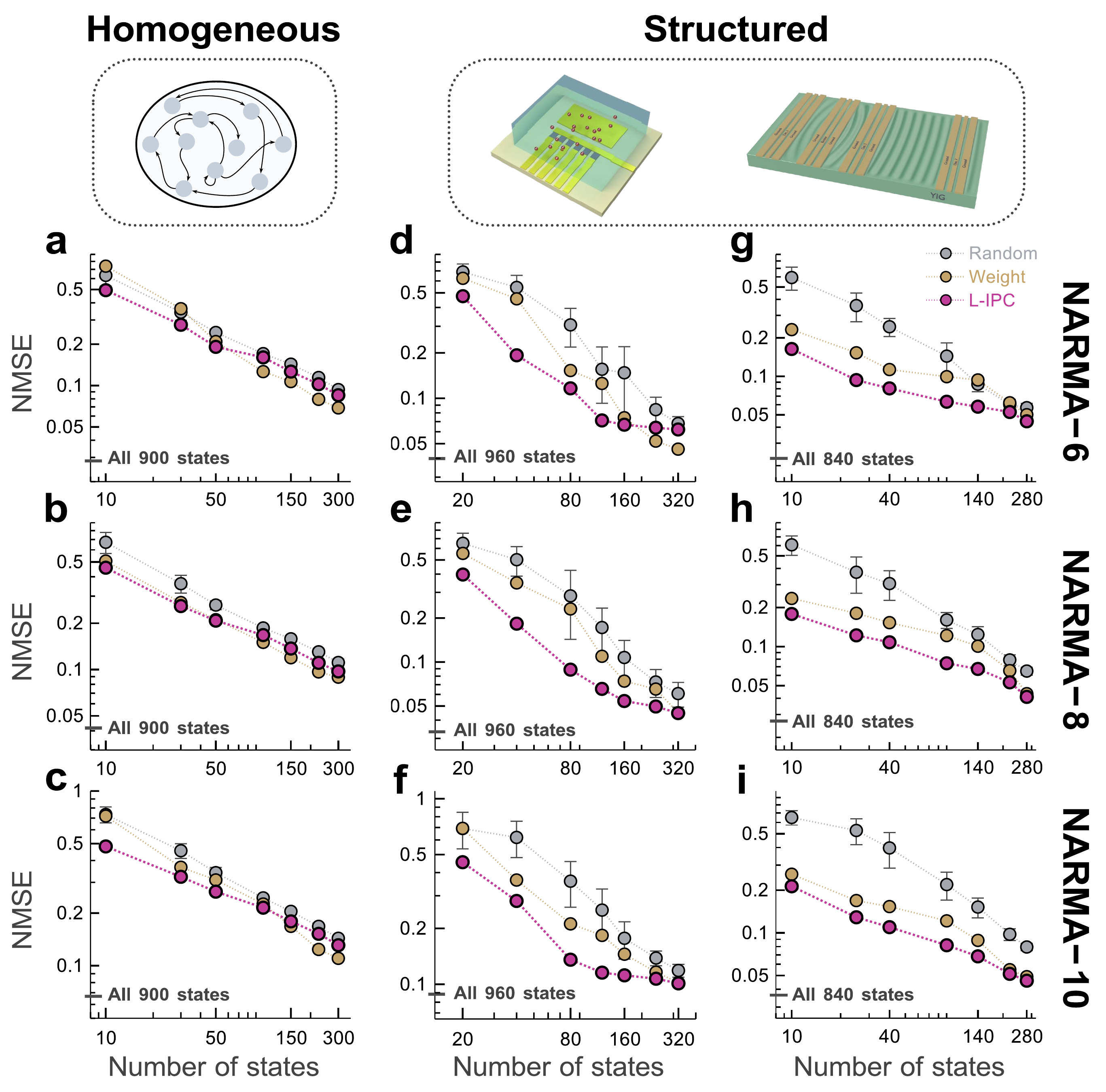}
\caption{\textbf{\textbar\ State reduction in computationally homogeneous and structured systems.} NMSE plotted against the number of selected states, for the computationally homogeneous ESN (\textbf{a}--\textbf{c}, $I = 900$) and the two computationally structured systems, the IGR (\textbf{d}--\textbf{f}, $I = 960$) and the SWR (\textbf{g}--\textbf{i}, $I = 840$), each evaluated on three tasks of NARMA-$6$
(top row), NARMA-$8$ (middle row) and NARMA-$10$ (bottom row). Three selection approaches are compared: L-IPC-guided (magenta), readout-weight-based (ochre) and random (gray). The dotted line gives the NMSE obtained using all available states. Error bars on the random selection give the standard deviation over $20$ independent draws.}
\label{fig:pruning}
\end{figure}

To examine the generality of these results, we apply the same three selection approaches to all three systems across a range of subset sizes for NARMA-$6$, $8$ and $10$ (Fig.~\ref{fig:pruning}). For this comparison, we also expand the ESN to $I = 900$ states by increasing the network size and the SWR to $I = 840$ states by using seven detectors (Methods). Their L-IPC profiles and selected states are shown in Supplementary Figs.~6 and~7, respectively. In the computationally homogeneous ESN, all three approaches yield comparable performance across the examined subset sizes (Fig.~\ref{fig:pruning}a--c), with L-IPC-guided selection retaining a slight advantage at smaller subsets and readout-weight-based selection slightly ahead at larger ones. The variation across random selections is also negligible, as reflected by the small error bars, indicating that different subsets offer similar computational capabilities. Thus, the choice of states has little influence on performance in such a homogeneous system.

In contrast, in the two structured systems, L-IPC-guided selection yields the lowest NMSE across nearly the entire range of subset sizes (Fig.~\ref{fig:pruning}d--i), with its advantage most pronounced for smaller subsets. The variation across random selections is also substantially larger, highlighting the importance of which states are retained. As the subset size increases, the performance of L-IPC-guided and readout-weight-based selection converges, with both approaching the full-state performance. Notably, retaining only one-third of the available states ($300$, $320$ and $280$ for the ESN, IGR and SWR, respectively) increases the NMSE by only about $20\%$ and $50\%$ relative to the full-state values for the IGR and SWR, whereas for the ESN it increases by a factor of two to three. These results demonstrate that structured computation is what makes system size reduction worthwhile and L-IPC-guided state selection effective.

\section*{Discussion}\label{sec:conclusion}

We have introduced L-IPC as a locally resolved metric that systematically quantifies the computational properties of individual dynamical states. Applying L-IPC to three systems with comparable global capacities revealed markedly different computational structures that reflect each system's underlying dynamics and remain hidden from a global view. We further demonstrated that these local profiles can guide task-adaptive state selection, outperforming readout-weight-based and random selection particularly in structured systems at small state numbers. Together, these results open a route toward interpretable neuromorphic computing and system miniaturization.

The state selection demonstrated here addresses a bottleneck of neuromorphic hardware. Device size, complexity and power consumption are directly linked to the number of states, as every state used must be physically read out, typically through a measurement channel followed by digitization~\cite{tanaka2019recent,yan2024emerging,kurebayashi2026metrics}. Reducing the readout size without substantially degrading performance is therefore important for bringing neuromorphic devices into practical use, particularly in edge and physical AI applications, where computational performance and efficiency are demanded together. A sufficiently small readout can, for example, make on-device or online learning feasible on a microcontroller. Although the selection criterion used here was rather naive, the L-IPC decomposition of each state provides a natural foundation for systematic pruning algorithms, and potentially for general training rules that act on the dynamics themselves to realize compact physical neural networks~\cite{momeni2025training}.

Importantly, computational irrelevance is a quantity that has not previously been available in the characterization of neuromorphic systems, for which it has conventionally been indistinguishable from redundancy. Nor is it resolved by a recently proposed state-resolved nonlinearity metric, which cannot separate nonlinear computation from irrelevant contributions~\cite{love2023spatial}. The distinction matters, as it diagnoses whether a device is limited by insufficient dynamical richness or by input-unrelated dynamics such as instability or noise (Supplementary Fig.~1). It also bears on practice, since we found that L-IPC-guided selection in the SWR loses its advantage when high-irrelevance states are not excluded first (Supplementary Fig.~8). Beyond these, L-IPC could help locate the edge of chaos, where computation is often optimal~\cite{legenstein2007edge}, since the irrelevance of states begins to rise as the dynamics destabilize (Supplementary Fig.~2). Furthermore, the L-IPC profiles show which states carry the richest dynamics and which are most robust to noise, indicating where additional resources would be best placed, such as the channels that would benefit most from further input delays. In this sense, L-IPC informs not only how well a device computes but how its dynamics and readout might be engineered to compute better, making it a natural candidate for post-fabrication screening of neuromorphic devices, analogous to the performance-based binning used in semiconductor manufacturing.

More broadly, because L-IPC characterizes each state on its own, it is not restricted to the reservoir-readout structure to which IPC is conventionally applied. It can therefore extend to individual units at any intermediate level of a larger architecture, from memristive spiking networks~\cite{dang2026spiking} to deep physical neural networks~\cite{wright2022deep}. The same construction can be readily extended to other input statistics and even to time-varying systems using time-dependent bases~\cite{PhysRevResearch.3.043135}, yielding a temporal L-IPC that tracks how computational roles change. Replacing the delays of a single input with multiple simultaneous inputs likewise brings the framework to multi-input memoryless neuromorphic systems, such as feedforward photonic~\cite{shen2017deep} or in-memory computing arrays~\cite{sebastian2020memory}. In practice, accurate coefficient estimation requires sufficient data and appropriate truncation of the expansion. Overall, L-IPC connects physical dynamics with computational function, thereby shifting the question from how much a system computes to where within it that computation resides, and offering a physically grounded basis for engineering neuromorphic devices toward their full capabilities.

\section*{Methods}\label{sec:method}

\subsection*{Computation of L-IPC}

The L-IPC profile of each dynamical state was computed using a Python package developed by the authors (see Code availability). A stochastic input signal $\{u(n)\}_{n=1}^{N}$ following a uniform distribution $\mathcal{U}(0, 0.5)$ with total length $N = 20000$ was used for all systems, including ESNs and the two physical reservoir platforms. The orthonormal basis functionals $\{\mathcal{Z}_{\bm{\upbeta}}[u]\}_{\bm{\upbeta}}$ were accordingly constructed as products of normalized Legendre polynomials. An initial transient of $n_0 = 1000$ time steps was discarded to eliminate initialization effects. The expansion coefficients $c^i_{\bm{\upbeta}}$ were estimated as the time-averaged inner product
\begin{equation}
    c^i_{\bm{\upbeta}} \approx \frac{1}{N - n_0} \sum_{n=n_0+1}^{N} \tilde{x}^i(n)\, \mathcal{Z}_{\bm{\upbeta}}[u](n).
    \label{eq:coeff_est}
\end{equation}
The Volterra expansion was truncated at maximum nonlinearity order $p_\mathrm{max} = 4$ and maximum memory step $k_\mathrm{max} = 15$, chosen by confirming that further increases in either parameter did not significantly change the results.

\subsection*{Echo state network setup}

We simulated an ESN with $I = 120$ states evolving according to
\begin{equation}
    \mathbf{x}(n+1) = f(\alpha \mathbf{W}_\mathrm{in} u(n) + \mathbf{W}_\mathrm{re} \mathbf{x}(n)),
    \label{eq:esn_eq}
\end{equation}
where $\mathbf{x}(n) = [x^1(n),\ldots,x^I(n)]^\top \in \mathbb{R}^{I}$ is the collection of $I$ states, $\alpha$ is the input scaling, $\mathbf{W}_\mathrm{in} \in \mathbb{R}^{I \times 1}$ is the input weight matrix, $\mathbf{W}_\mathrm{re} \in \mathbb{R}^{I \times I}$ is the recurrent weight matrix, and $f$ is the nonlinear activation function, taken here as the hyperbolic tangent. Both $\mathbf{W}_\mathrm{in}$ and $\mathbf{W}_\mathrm{re}$ were sampled independently from $\mathcal{U}(-1, 1)$ using a fixed random seed across all parameter settings, and $\mathbf{W}_\mathrm{re}$ was then rescaled to the target spectral radius $\rho(\mathbf{W}_\mathrm{re})$. For the ESN shown in Fig.~\ref{fig:interpretation}, we set $\alpha = 1$ and $\rho(\mathbf{W}_\mathrm{re}) = 1.3$. For the validation in Supplementary Note~1, we further varied $\alpha$ and $\rho(\mathbf{W}_\mathrm{re})$ independently to examine their effects on the L-IPC profiles. For the state selection in Fig.~\ref{fig:pruning}, we increased the network size to $I = 900$, with $\alpha = 0.5$ and $\rho(\mathbf{W}_\mathrm{re}) = 1.2$ fixed. 

\subsection*{Ion-gating reservoir setup}

A side-gated ion-gel/graphene transistor array was fabricated on a Si substrate covered by a 300-nm-thick SiO$_2$ layer. Monolayer graphene grown by chemical vapor deposition and transferred onto the substrate (Graphenea, Spain) was patterned by dry etching into six nominally identical channels, each 40~\textmu m wide and 10~\textmu m long. The six channels shared a common source electrode and a common side-gate electrode and were connected to independently addressable drain electrodes. The electrodes were patterned by photolithography, followed by electron-beam evaporation of Cr (10~nm) and Au (50~nm) and lift-off. An approximately 200-\textmu m-thick ion-gel consisting of 1-ethyl-3-methylimidazolium bis(trifluoromethanesulfonyl)imide (EMIm-TFSI), poly(dimethylaminoethyl methacrylate) (PDMEMA), and $N,N,N',N'$-tetra(trifluoromethanesulfonyl)-dodecane-1,12-diamine (C12TFSA) was prepared following Ref.~\cite{nishioka2025two} and placed over the active region, covering both the graphene channels and the side-gate electrode. The assembled device was dried overnight in a vacuum chamber.

Electrical measurements were performed at room temperature in a vacuum chamber using the pulse measurement units of a 4200A-SCS semiconductor parameter analyzer (Keithley Instruments, USA). The device chip was mounted on an IC package and wire-bonded to the package terminals. Each discrete-time input $u(n)\in[0,0.5]$ was linearly mapped to a voltage-pulse amplitude between 0 and 2~V, with a duty cycle of 50\%. A constant voltage of 0.5~V was applied to drain~1, while the common gate and drains~2--6 were driven by $u(n)$, $u(n-2)$, $u(n-4)$, $u(n-6)$, $u(n-8)$, and $u(n-10)$, respectively. For the IGR shown in Fig.~\ref{fig:interpretation}, the pulse interval was set to $80$~\textmu s. The gate, source and six drain currents were simultaneously recorded, yielding eight current responses. From each response, $15$ points (virtual nodes) were extracted by evenly spaced temporal sampling within a single pulse interval, resulting in $120$ states in total. For the state selection in Fig.~\ref{fig:pruning}, we expanded the state set by operating the IGR at eight pulse intervals ($80$, $60$, $40$, $20$, $10$, $8$, $6$ and $4$~\textmu s), each driven by the same input sequence, giving $960$ states in total.

\subsection*{Spin-wave reservoir setup}

The SWR comprises a double-side-polished, $(111)$-oriented YIG single crystal measuring $7 \times 7 \times 0.5$~mm$^3$, grown by the flux method and supplied by TwoLeads Corporation. Ten CPWs were patterned on the YIG surface by laser lithography and formed by electron-beam evaporation of a Ti ($10$~nm)/Au ($90$~nm) bilayer, each consisting of a $10$-\textmu m-wide signal line and two $20$-\textmu m-wide ground lines separated by $10$-\textmu m gaps. The center-to-center distances between adjacent signal lines, from left to right, were $75$, $75$, $85$, $75$, $85$, $75$, $85$, $75$ and $80$~\textmu m. The second CPW from the left was used as the exciter, and for the SWR shown in Fig.~\ref{fig:interpretation} the tenth CPW was used as the detector.

During measurements, an out-of-plane bias magnetic field of $190.3$~mT was applied. The discrete-time input $u(n)$ was encoded into voltage pulses with amplitudes of $0$--$300$~mV, a rise/fall time of $0.375$~ns, a flat-top duration of $0.375$~ns and a start-to-start interval of $2$~ns, following Ref.~\cite{chen2026spectral}. Pulses were generated by an arbitrary waveform generator (Keysight M8915A) at a sampling rate of $16$~GHz and applied to the exciter through a $24.5$-dB-gain power amplifier (ZVA-0.5W303G). The detector responses were amplified by $34$-dB-gain amplifiers (PE15A3269) and recorded with an oscilloscope (Tektronix MSO68B) at a sampling rate of $12.5$~GHz, with 500-shot averaging for noise suppression.

States were extracted numerically in MATLAB by applying second-order Butterworth band-pass filters with a $3$-dB passband width of $0.1$~GHz. The $120$ filter center frequencies were spaced by $0.025$~GHz from $0.125$ to $3.1$~GHz. Envelope detection was then performed by taking the root-mean-square value of each filtered signal over each $2$-ns input period, yielding $120$ states in total (spectral nodes). For the state selection in Fig.~\ref{fig:pruning}, we expanded the state set by using seven CPWs as detectors (the first, third, fifth, sixth, seventh, eighth and tenth from the left), giving $840$ states in total.

\subsection*{Computation of global IPC}

The global IPC in Fig.~\ref{fig:interpretation}d--f was computed following Ref.~\cite{dambre2012information}, using the same input signal $u(n)$ and orthonormal basis $\{\mathcal{Z}_{\bm{\upbeta}}[u]\}_{\bm{\upbeta}}$ as for the L-IPC, with truncation at $p_\mathrm{max} = 4$, $k_\mathrm{max} = 15$. For each basis functional, the capacity
\begin{equation}
    C[\mathcal{Z}_{\bm{\upbeta}}] = 1 - \frac{\min_{\mathbf{w}} \sum_{n = n_0+1}^{N} \big( \mathcal{Z}_{\bm{\upbeta}}[u](n) - \mathbf{w}^{\top}\tilde{\mathbf{x}}(n) \big)^{2}}{\sum_{n = n_0+1}^{N} \mathcal{Z}_{\bm{\upbeta}}^{2}[u](n)}
    \label{eq:global_ipc}
\end{equation}
was obtained by linear regression of the target onto all $I$ normalized states $\tilde{\mathbf{x}}(n) = \big(\tilde{x}^1(n), \ldots, \tilde{x}^I(n)\big)^{\top}$, and the total IPC was taken as $\sum_{\bm{\upbeta}} C[\mathcal{Z}_{\bm{\upbeta}}]$. Capacities below twice the maximum capacity obtained from $20$ surrogate datasets, generated by randomly permuting the target sequences in time, were set to zero to prevent overestimation of the total IPC due to finite data length.

\subsection*{NARMA tasks and readout evaluation}

The NARMA-$l$ task is to predict the output of the discrete-time recursive system
\begin{equation}
y(n) = 0.3\, y(n-1) + 0.05\, y(n-1) \sum_{k=1}^{l} y(n-k) + 1.5\, u(n-1)\, u(n-l) + 0.1,
\label{eq:narma}
\end{equation}
driven by an input sequence $u(n)$ drawn independently from a uniform distribution $\mathcal{U}(0, 0.5)$. The input enters through the second-order product $u(n-1)\,u(n-l)$, and the recursion carries the dependence back to delay $l$, so that NARMA-$l$ predominantly requires second-order nonlinearity together with memory at delay step $l$~\cite{PhysRevResearch.3.043135}. We therefore take $\mathcal{N}_2^i$ and $\mathcal{M}_l^i$ as the task-relevant nonlinearity and memory components for NARMA-$l$.

The reservoir output is obtained as a linear combination of the dynamical states,
\begin{equation}
\hat{y}(n) = \sum_{i \in S} w_i\, \tilde{x}^i(n) + w_0,
\end{equation}
where $S$ denotes the set of selected states and $w_0$ is a bias weight. The states $\tilde{x}^i(n)$ were normalized so that the readout weights are directly comparable across states in the weight-based selection. The weights were determined by least-squares regression, $\mathbf{w} = \tilde{X}^{+}\mathbf{y}$, where $\mathbf{w}$ collects the readout weights $w_i$ and the bias $w_0$, $\tilde{X}^{+}$ is the Moore--Penrose pseudoinverse of the normalized state matrix $\tilde{X}$ augmented with a constant column, and $\mathbf{y}$ is the vector of target values. Sequences of $N_\mathrm{total} = 20000$ steps were used, of which the first $1000$ were discarded as washout, $12000$ for training and the remaining $7000$ for testing. We evaluated task performance by calculating the NMSE between the predicted output $\hat{y}(n)$ and the target output $y(n)$ on the test set,
\begin{equation}
\mathrm{NMSE} = \frac{\sum_n \left(\hat{y}(n) - y(n)\right)^2}{\sum_n \left(y(n) - \bar{y}\right)^2},
\end{equation}
where $\bar{y}$ is the mean of the target output.

\subsection*{State selection procedures}

For each task, a subset $S$ of the $I$ available states was selected, with $I = 900$, $960$ and $840$ for the ESN, IGR and SWR, respectively (Fig.~\ref{fig:pruning}). Three selection procedures were compared, and in each case, the readout was retrained on the selected subset alone.

In L-IPC-guided selection, we first discarded all states with computational irrelevance $\eta^i > 0.1$. Among the remainder, we retained the $2|S|$ states with the largest task-relevant memory $\mathcal{M}_l^i$ for NARMA-$l$, where $|S|$ is the target system size. Finally, the states with the largest $\mathcal{N}_2^i$ among these were taken until the target size was reached. No reference was made to prediction performance at any stage of the selection.

In readout-weight-based selection, a readout was first trained on all $I$ normalized states $\{\tilde{x}^i(n)\}_{i=1}^{I}$. States were then ranked by the absolute value of their readout weight $|w_i|$, and the highest-ranked states were taken until the target size was reached.

In random selection, states were drawn uniformly from the $I$ available states. The procedure was repeated independently $20$ times, and the mean NMSE and its standard deviation are reported.

\bmhead{Data availability} The data supporting the findings of this study are available within the article and its supplementary information files. The raw measurement data are available from the corresponding author on reasonable request.

\bmhead{Code availability} The L-IPC analysis package developed for this study is provided as Supplementary Software for peer review and will be made openly available on GitHub, with an archived release on Zenodo, upon publication.

\bmhead{Acknowledgements} This work was supported in part by Innovative Science and Technology Initiative for Security Grant Number JPJ004596, ATLA, Japan, and in part by ``Advanced Research Infrastructure for Materials and Nanotechnology in Japan (ARIM)’’ of the Ministry of Education, Culture, Sports, Science and Technology (MEXT), Proposal Numbers JPMXP1226NM5256, 26NM5258 and 26NM5259. D.N. was supported by JSPS KAKENHI Grant Numbers JP24KJ0299 (Grant-in-Aid for JSPS Fellows) and JP25K17941. R.I. was supported by JSPS KAKENHI Grant Number JP24K01334.

\bmhead{Author contributions} J.C. and Y.S. conceived the study. J.C., Y.S. and R.I. developed the L-IPC framework. J.C. and S.H. wrote the L-IPC software package. J.C. performed the ESN simulations. D.N. and H.K. fabricated the IGR device, and D.N. performed the IGR measurements. J.C. and S.H. fabricated the SWR device, and J.C. performed the SWR measurements. J.C. analyzed the data and prepared the figures. J.C. wrote the original draft. J.C., R.I., A.H. and T.T. reviewed and edited the manuscript. R.I., A.H. and T.T. supervised the project. T.T. acquired funding. All authors discussed the results and commented on the manuscript.

\bmhead{Competing interests}

The authors declare no competing interests.


\input Main.bbl

\clearpage
\includepdf[pages=-]{supplement.pdf}

\end{document}